\documentclass[aps,prl,twocolumn,superscriptaddress,oneside,floatfix,amsmath,showpacs,amssymb]{revtex4}
\usepackage{graphicx}
\usepackage{dcolumn}
\usepackage{bm}

\begin{document}

\newcommand{\Y}{YBa$_2$Cu$_3$O$_{6.5}$}
\newcommand{\ie}{\textit{i.e.}}
\newcommand{\eg}{\textit{e.g.}}
\newcommand{\etal}{\textit{et al.}}

%%%%%%%%%%%%%%%%%%%%%%%%%%%% TITLE

\title{Multiple Quantum Oscillations in the de Haas van Alphen Spectra
of the Underdoped High Temperature Superconductor
YBa$_2$Cu$_3$O$_{6.5}$}

%%%%%%%%%%%%%%%%%%%%%%%%%%%% AUTHORS
\author{Alain~Audouard}
\affiliation{Laboratoire National des Champs Magn\'{e}tiques
Intenses (CNRS), Toulouse, France}

\author{Cyril~Jaudet}
\affiliation{Laboratoire National des Champs Magn\'{e}tiques
Intenses (CNRS), Toulouse, France}

\author{David~Vignolles}
\affiliation{Laboratoire National des Champs Magn\'{e}tiques
Intenses (CNRS), Toulouse, France}

\author{Ruixing~Liang}
\affiliation{Department of Physics and Astronomy, University of
British Columbia, Vancouver, Canada} \affiliation{Canadian
Institute for Advanced Research, Toronto, Canada}

\author{D.A.~Bonn}
\affiliation{Department of Physics and Astronomy, University of
British Columbia, Vancouver, Canada} \affiliation{Canadian
Institute for Advanced Research, Toronto, Canada}

\author{W.N.~Hardy}
\affiliation{Department of Physics and Astronomy, University of
British Columbia, Vancouver, Canada} \affiliation{Canadian
Institute for Advanced Research, Toronto, Canada}

\author{Louis~Taillefer}
\affiliation{Canadian Institute for Advanced Research, Toronto,
Canada} \affiliation{D\'epartement de physique \& RQMP,
Universit\'e de Sherbrooke, Sherbrooke, Canada}

\author{Cyril~Proust} \email{proust@lncmp.org}
\affiliation{Laboratoire National des Champs Magn\'{e}tiques
Intenses (CNRS), Toulouse, France} \affiliation{Canadian Institute
for Advanced Research, Toronto, Canada}

\date{\today}
%Version 4.0

%%%%%%%%%%%%%%%%%%%%%%%%%%%% ABSTRACT

\begin{abstract}
By improving the experimental conditions and extensive data
accumulation, we have achieved very high-precision in the
measurements of the de Haas-van Alphen effect in the underdoped
high-temperature superconductor YBa$_{2}$Cu$_{3}$O$_{6.5}$. We
find that the main oscillation, so far believed to be
single-frequency, is composed of three closely spaced frequencies.
We attribute this to bilayer splitting and warping of a single
quasi-2D Fermi surface, indicating that \emph{c}-axis coherence is
restored at low temperature in underdoped cuprates. Our results do
not support the existence of a larger frequency of the order of
1650~T reported recently in the same compound [S.E. Sebastian {\it
et al}., Nature  {\bf 454}, 200 (2008)].

\end{abstract}


\maketitle

%%%%%%%%%%%%%%%%%%%%%%%%%%%% INTRODUCTION

The first unambiguous observation of quantum oscillations (QO) in
high temperature superconductors (HTSC) \cite{Doiron07} has raised
much interest concerning the origin of their Fermi surface (FS).
On the overdoped side, recent quantum oscillations measurements in
Tl$_2$Ba$_2$CuO$_{6+\delta}$ \cite{Vignolle08} (Tl-2201) have
confirmed that the Fermi surface consists of a single large
cylinder covering 65$\%$ of the first Brillouin zone (FBZ) as
shown by angular magnetoresistance oscillations (AMRO)
\cite{Hussey03} or angle-resolved photoemission spectroscopy
(ARPES) \cite{Plate05}, in agreement with Hall effect data
\cite{Mackenzie96} and band structure calculations based on
local-density approximation (LDA) \cite{Singh92,Andersen95}. This
result demonstrates that QO probe directly the FS originating from
the CuO$_2$ planes and that the
in-field QO agree with zero-field measurements.\\
\indent On the underdoped side, both Shubnikov-de Haas (SdH)
\cite{Doiron07,Bangura08,Yelland08} and de Haas-van Alphen (dHvA)
\cite{Jaudet08,Sebastian08} oscillations have been observed in
underdoped YBa$_2$Cu$_3$O$_{y}$ (YBCO) pointing to a FS
dramatically different from that of overdoped Tl-2201. Indeed,
provided that quantum oscillations are interpreted as the
consequence of quantization of closed orbits in a magnetic field
\cite{Shoenberg}, it can be inferred that the FS consists of small
pockets covering only about 2~$\%$ of the FBZ
\cite{Doiron07,Bangura08,Yelland08}. One fundamental question for
our understanding of high temperature superconductors is what
causes the dramatic change in FS topology from overdoped to
underdoped regimes\cite{Chakravarty08b,Jaudet09, Taillefer09}?
Based on the observation of a negative Hall effect at low
temperature in underdoped YBCO \cite{LeBoeuf07}, the presence of
an electron pocket in the Fermi surface has been suggested. It
would naturally arise from a reconstruction of the large hole FS
calculated from band structure. Different mechanisms have been
proposed for this FS reconstruction, involving \emph{d}-DW order
(a scenario based on a hidden broken symmetry of
d$_{x^2-y^2}$-type) \cite{Chakravarty08}, incommensurate
antiferromagnetic order \cite{Sebastian08,Harrison07}, stripe
order \cite{Millis07} or field induced spin density wave state
\cite{Chen08,Sachdev09}. Each scenario predicts a different FS
topology and the observation of an additional frequency would
provide a stringent test. In particular, recent dHvA measurements
\cite{Sebastian08} (performed in the same compound grown at the
University of British Columbia as in ref.~\cite{Jaudet08} and the
present study) seems to indicate the presence of a larger pocket,
which would impose a strong constraint on the ordering wave vector
for the reconstruction.

\indent In order to determine whether there are other closed
sections of the FS, something of fundamental importance for
clarifying the FS of HTSC in the pseudogap phase, we have
performed high-precision measurements of the de Haas-van Alphen
effect in underdoped YBa$_2$Cu$_3$O$_{y}$ ($y$=6.51 and $y$=6.54).
We found no evidence of an oscillatory component of high frequency
as reported in ref~\cite{Sebastian08} but the sensitivity of the
measurements allow the uncovering of some important details for
the electronic structure of underdoped YBCO. In addition to the
main frequency $F_1$=540~T, two closely spaced satellites
$F_2$=450~T and $F_3$=630~T have been resolved. We attribute these
frequencies to a bilayer effect and warping of the FS, which
proves that quasiparticle motion along the inter-plane direction
is coherent at low temperature in the underdoped regime.

%%%%%%%   EXPERIMENTAL
In order to achieve high-precision measurements of the de Haas-van
Alphen effect, we have averaged several pulses (up to 10) for the
same experimental conditions, as described in more detail in
[\onlinecite{Jaudet08}]. We have measured two single crystals:
YBa$_2$Cu$_3$O$_{6.51}$ ($T_c$=57.5~K) and YBa$_2$Cu$_3$O$_{6.54}$
($T_c$=60~K), flux-grown in a non-reactive BaZrO$_3$ crucible. The
dopant oxygen atoms were ordered into the ortho-II superstructure
of alternating full and empty CuO chains \cite{Liang00}. Typical
dimensions of the samples are 140$\times$140$\times$40~$\mu$m$^3$.
A careful study of the amplitude of the oscillations between the
up (50~ms) and down (250~ms) field sweeps shows that the sample
temperature changes very little during the pulse. Given an
effective mass of $m^*$=1.8~m$_0$ \cite{Jaudet08}, the upper bound
for the temperature difference is about 0.2~K at $T$=0.7~K, which
excludes any large heating effects during the pulse.

\begin{figure}                                                  %Fig. torque 6.51 et 6.54
\centering
\includegraphics[width=0.85\linewidth,angle=0,clip]{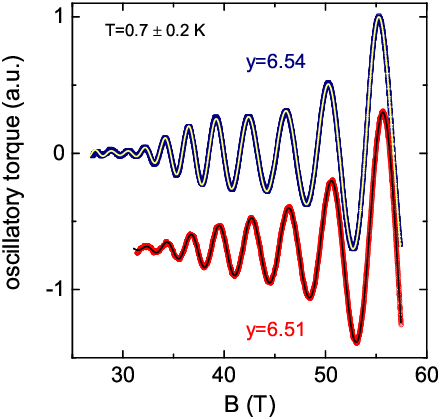}
\caption{ (color on line) Oscillatory part of the torque signal
for underdoped YBa$_2$Cu$_3$O$_{6.54}$ (upper blue line) and
YBa$_2$Cu$_3$O$_{6.51}$ (lower red line). Solid lines are best
fits of Eq.~\ref{eq:LK} to the data. Four Fourier components are
involved in the fit (see text).}\label{fig1}
\end{figure}

Fig.~\ref{fig1} displays the field dependence of the oscillatory
torque in two ortho-II samples, obtained after averaging up to 10
pulses during the down-field sweep at $T$=0.7$\pm$0.2~K. A
polynomial monotonic background has been subtracted from the raw
data. While the amplitude of the oscillations should grow
exponentially with magnetic field for a single frequency, a
modulation of the amplitude of the oscillation (beating effect)
can be noticed in both samples, which is the signature of the
existence of more than one frequency in the oscillatory spectrum.
Solid lines in Fig.~\ref{fig1} are fits of the Lifshitz-Kosevich
(LK) theory, assuming that four frequencies are involved in the
data, as discussed in more detail below.

In the framework of the LK theory, the field-dependent oscillatory
part of the torque, for a multiband 2D Fermi surface, can be
written:
$\tau_{osc}=B\sum_{i}A_{i}\sin[2\pi(\frac{F_{i}}{B}-\gamma_{i})]$
where $F_{i}$ is the oscillation frequency linked to a given orbit
and $\gamma_{i}$ is a phase factor. Neglecting magnetic breakdown
and spin damping contributions, the amplitude of a given Fourier
component A$_{i}$ $\propto$ R$_{Ti}$R$_{Di}$ where R$_{Ti}$ =
$\alpha$Tm${_i^*}$/Bsinh[$\alpha$Tm${_i^*}$/B] and R$_{Di}$ =
exp[-$\alpha$T$_{Di}$m${_i^*}$/B] are the thermal and Dingle
damping factors, respectively. Here,
$\alpha=2\pi^2k_Bm_0/e\hbar$=14.69~Tesla/Kelvin and $T_{Di}$ is
the Dingle temperature \cite{Shoenberg}. In order to restrict the
number of free parameters in the data analysis, we have considered
the asymptotic form:
\begin{equation}
\label{eq:LK}
{\tau_{osc}}=\sum_{i}C_i\exp(-\frac{B_i}{B})\sin[2\pi(\frac{F_{i}}{B}-\gamma_{i})]
\end{equation}
where B$_i$ = $\alpha$(T+T$_{Di}$)m${_i^*}$. Strictly speaking,
the latter assumption is only valid in the high $T/B$ range and
only affects the amplitude of the signal. The analysis is based on
discrete Fourier transforms (DFT) using Blackman windows, which
are known to avoid secondary lobes.

\begin{figure}                                                        % Figure fit 6.54
\centering
\includegraphics[width=0.85\linewidth,angle=0,clip]{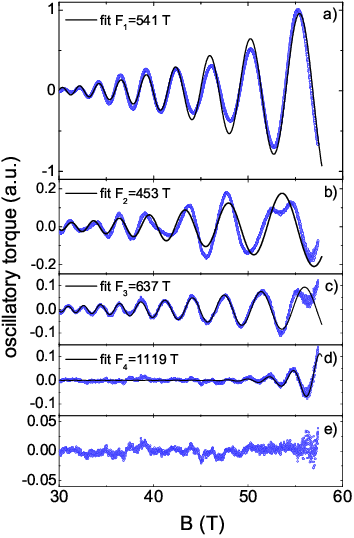}
\caption{(color on line) (a) Oscillatory part of the torque signal
for YBCO$_{6.54}$ (symbols). Black solid line is the best fit of
Eq.~\ref{eq:LK} to the data with only the main frequency
$F_1$=541~T. (b) Residual (fit-data) of the previous fitting
(symbols). The line shows the best fit of Eq.~\ref{eq:LK} to the
data with two frequencies $F_1$ and $F_2$=453~T. The two other
panels show the same procedure by taking into account (c) a third
frequency $F_3$=637~T  and (d) a fourth frequency $F_4$=1119~T.
(e) Residual of the final fit.}\label{fig2}
\end{figure}

Fig.~\ref{fig2}(a) displays the field dependence of the
oscillatory torque of YBa$_2$Cu$_3$O$_{6.54}$ (symbols). The solid
line is a fit to the data assuming only one frequency $F_1$=541~T
(solid line), which cannot match the dHvA data over a wide field
range. This shows that the data \emph{cannot} be accounted for by
only one Fourier component. A large residual (i.e. the difference
between the fit and data) can be observed in Fig.~\ref{fig2}(b)
(symbols). The oscillatory spectrum still looks complicated, which
means that more than one frequency is involved. DFT on this
residual reveals the presence of two frequencies close to $F_1$.
The black line in Fig.~\ref{fig2}(b) is a fit to the data with a
second Fourier component $F_2$=453~T, whose residual is now
displayed in Fig.~\ref{fig2}(c) (symbols). The solid line in
Fig.~\ref{fig2}(c) is a fit with a third frequency $F_3$=637~T.
Finally, the residual in Fig.~\ref{fig2}(d) still contains a small
component at $F_4$=1119~T which we assume is the second harmonic
of $F_1$. The final residual is displayed in Fig.~\ref{fig2}(e).
The same kind of analysis performed on YBa$_2$Cu$_3$O$_{6.51}$
gives similar results. The averaged frequencies deduced from the
fits of Eq.~\ref{eq:LK} to the oscillatory torque of the two
samples are $F_1$=540$\pm$15~T, $F_2$=450$\pm$15~T,
$F_3$=630$\pm$40~T and $F_4$=1130$\pm$20~T.

\begin{figure}                                                        % Figure TF
\centering
\includegraphics[width=0.85\linewidth,angle=0,clip]{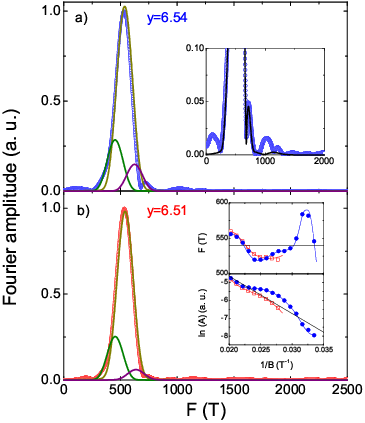}
\caption{(color on line) Fourier analysis of the oscillatory
torque (see Fig.~\ref{fig1}) for a) YBa$_2$Cu$_3$O$_{6.54}$ and b)
YBa$_2$Cu$_3$O$_{6.51}$ in the field range 31.4 to 57.4~T. Solid
lines show the contribution of $F_1$ (yellow), $F_2$ (green) and
$F_3$ (purple), respectively. The top inset shows a zoom of the
DFT of the data (symbols) and of the fit including four
frequencies (black line). The bottom inset shows the field
dependence of the main peak of the DFT and of the oscillation
amplitude. Symbols and lines correspond to DFT of the data and DFT
of the fit including four frequencies, respectively.}\label{fig3}
\end{figure}

The corresponding Fourier analysis of the data for
YBa$_2$Cu$_3$O$_{6.54}$ and YBa$_2$Cu$_3$O$_{6.51}$ are displayed
in symbols in Fig.~\ref{fig3}(a) and Fig.~\ref{fig3}(b),
respectively. A broad peak with a maximum around 535 T is
observed, in agreement with \cite{Doiron07,Jaudet08}. The DFT of
the fit involving only the $F_1$ component (yellow lines in
Fig.~\ref{fig3}) cannot reproduce the tail and the small maxima on
either side of the broad peak. These features are further evidence
for the two additional frequencies $F_2$ (green lines in
Fig.~\ref{fig3}) and $F_3$ (purple lines in Fig.~\ref{fig3}),
respectively. The DFT of the fit including all the Fourier
components successfully reproduces these features, as shown in the
inset of Fig.~\ref{fig3}(a) which displays a zoom of the DFT of
the data (symbols) and of the fit (black line) for
YBa$_2$Cu$_3$O$_{6.54}$. The highest frequency detected in both
measurements is $F_4$. The field dependence of the main peak of
the DFT and of the oscillation amplitude ('Dingle plot') are
displayed in the inset of Fig.~\ref{fig3}(b). DFT of the data
(symbols) and DFT of the fit (solid lines) are in very good
agreement. The black lines show the behavior expected if only one
frequency was involved. The strongly non-monotonic
field-dependence of the frequency of the main peak and the
apparent non-LK behavior of its amplitude are a signature of
several Fourier components with frequencies in a narrow range.

Let us now discuss scenarios which may provide an explanation for
the beating effect observed in QO measurements. While it cannot be
ruled out that additional FS pockets account for the observed
oscillatory spectra, the most natural interpretation is to invoke
both a slight modulation of the FS sheet along the \emph{c}-axis
(warping) and a bilayer splitting effect, the latter being
intrinsic to YBa$_2$Cu$_3$O$_{y}$ which contains two closely
spaced CuO$_2$ planes (separated by 3.2~$\AA$). The bilayer
splitting arises from bonding-antibonding coupling of interplane
wave functions and gives rise to two CuO$_2$ plane-derived FS
pockets \cite{Andersen95}. In band structure calculations for
ortho II-\Y~in ref.~\cite{Carrington07}, the warping and bilayer
effects led naturally to four frequencies: two come from the
bilayer effect and each of them is further split due to warping of
the FS (providing that $c$-axis modulation leads to two different
extremal areas). Assuming that $F_4$ is a second harmonic of $F_1$
(but we can't rule out that $F_4$ is a second frequency with a
small amplitude corresponding to the hole pocket in a commensurate
FS reconstruction \cite{Chakravarty08}), the observation of three
frequencies, rather than the expected four, can be explained
either if two frequencies are too close to be resolved or if the
amplitude of one frequency is too weak.

There are two hopping terms to consider: $t_{\bot}^c$ comes from
the hopping through layers in between the CuO$_2$ planes and
$t_{\bot}^b$ represents the hopping between the planes of the
bilayer. In order to estimate $t_{\bot}^c$ which causes warping of
the FS along the \emph{c}-axis, we follow the calculation by
Yamaji \cite{Yamaji98} and assume the simplest form for warping.
The quasi-2D dispersion can be written as
$\epsilon=\hbar^2/2m^*(k_x^2+k_y^2)-2t_{\bot}^c cos(ck_z)$, where
$c$ is the \emph{c}-axis lattice parameter. The $c$-axis warping
leads to two extremal FS areas perpendicular to the magnetic field
and therefore to two close frequencies that yield a beating
effect. From geometrical considerations, the frequency difference
is given by $\Delta F=4m^*t_{\bot}^c/e\hbar$. Assuming that
$\Delta F$=90~T ($F_1$-$F_2$ or $F_3$-$F_1$), the $c$-axis hopping
term $t_{\bot}^c\simeq$1.3~meV, i.e. 15~K. This value is similar
to that deduced from the beating effect induced by the warping of
the quasi-2D $\beta$ FS sheet in the quasi-2D oxide
Sr$_2$RuO$_4$~\cite{Mackenzie03}.

In order to deduce a value for the bilayer hopping term
$t_{\bot}^b$, it is necessary to rely on band structure
calculations. A model which assumes a reconstruction of the FS can
account for the splitting of the frequency linked to the electron
pocket frequency into two frequencies $F$=570~T and $F$=450~T by
considering a bilayer coupling $t_{\bot}^b$=8~meV \cite{Dimov08}.
For the unreconstructed FS, band structure calculations predict
that $t_{\bot}^b\simeq$200~meV \cite{Andersen95,Carrington07}.
This striking difference in the value of $t_{\bot}^b$ is another
indication of the reconstruction of the large hole FS
\footnote{Note that in YBa$_2$Cu$_4$O$_{8}$, the same order of
discrepancy has been deduced from magnetotransport measurements
for the interchain coupling, which has been interpreted as virtual
hopping process \cite{Hussey98}.}.

If warping and bilayer effects are responsible for the observation
of several frequencies in underdoped \Y, it thus implies that
quasiparticle motion is coherent along the $c$-axis at low
temperature. While a coherent 3D-FS has been established by AMRO
experiments in overdoped Tl-2201 \cite{Hussey03}, a crossover in
the phase diagram between a coherent metal phase (overdoped) and
an incoherent one (underdoped) has been suggested in
Bi$_{2}$Sr$_{2}$CaCu$_2$O$_{8+\delta}$ \cite{Kaminski03}. Indeed,
both electrical \cite{Takenaka94, Ando96} and optical measurements
\cite{Tamasaku92, Basov99, Homes05} show that in-plane response
can be metallic whereas the out-of-plane response is
insulating-like. The present measurements show that at
sufficiently low temperature, when superconductivity is suppressed
by a strong magnetic field, the coherence along the c-axis is
restored in underdoped and clean material, i.e. it establishes the
existence of a coherent three-dimensional FS, concomitant with the
observation of QO.

%%%% ARPES - 2nd frequency
We end by discussing some open questions raised by the QO in
underdoped YBCO. The first is the apparent contradiction with
ARPES, which shows the destruction of the Fermi surface in the
antinodal regions, producing a set of disconnected Fermi arcs
\cite{Norman98,Hossain08}. Assuming that ARPES detects only one
side of a hole pocket at ($\pi$/2, $\pi$/2) \cite{Chakravarty03,
Harrison07,Meng09} is not sufficient to explain any reconstruction
scenario. Indeed, there is no evidence so far in ARPES data of
other pockets or regions of the FS where coherent quasiparticle
exist, except around ($\pi$/2, $\pi$/2). In a stripe scenario
\cite{Millis07}, the FS contains quasi-1D hole-like Fermi
surfaces, for which there is no evidence in ARPES data. One
proposed scenario to reconcile ARPES and QO measurements is a
field induced state\cite{Chen08,Sachdev09}, as suggested by recent
neutron scattering experiment \cite{Haug09}. The second open
question concerns the larger frequency ($F_{\beta}\simeq$1650~T),
which has been detected with a very small amplitude in underdoped
YBCO ortho-II \cite{Sebastian08}. The observation of this
frequency led the authors of ref.~\onlinecite{Sebastian08} to
suggest a FS reconstruction of the large hole FS into several
pockets with different mobilities. In the present study, we were
not able to detect this high frequency within a noise level lower
than 0.5$\%$ of the amplitude of $F_1$, while the amplitude of
this high frequency ($F_{\beta}$) is reported to be 3$\%$ of the
main frequency ($F_1$) in ref.~\cite{Sebastian08}. Further
experiments on different samples should help clarify this apparent
contradiction.

In summary, we have achieved very high-precision in the
measurements of the de Haas-van Alphen effect in underdoped
YBa$_{2}$Cu$_{3}$O$_{6.5}$. We found no evidence of a larger
pocket as claimed in ref~\cite{Sebastian08}. A fitting procedure
based on the Lifshitz-Kosevich theory has revealed that the main
oscillation is in fact composed of at least three closely spaced
frequencies, which we attribute to bilayer splitting and warping
of the FS along the \emph{c}-axis. This interpretation implies
that quasiparticle transport perpendicular to the CuO$_2$ planes
is coherent in the underdoped regime at low temperature.

%%%%%%% ACKNOWLEDGEMENTS  %%%%%%%%%%%%%%%%

We thank A. Carrington, S. Chakravarty, N. Hussey, M. Norman, G.
Rikken, S.E. Sebastian, B. Vignolle and J. Zaanen for useful
discussions. We acknowledge support from EuroMagNET under EC
contract 506239, the ANR projects ICENET and DELICE, the Canadian
Institute for Advanced Research and funding from NSERC, FQRNT,
EPSRC, and a Canada Research Chair.

\end{document}